\documentstyle[11pt,aasms]{article}

\tighten
\slugcomment{Submitted for publication in 
{\it The Astrophysical Journal Letters}}

\begin{document}



\def\pmb#1{\setbox0=\hbox{#1}%
  \kern0.00em\copy0\kern-\wd0
  \kern0.03em\copy0\kern-\wd0
  \kern0.00em\raise.04em\copy0\kern-\wd0
  \kern0.03em\raise.04em\copy0\kern-\wd0\box0 }

\def\pp{\parshape=2 -0.25truein 6.75truein 0.5truein 6truein}

\def\ref #1;#2;#3;#4;#5{\par\pp #1 #2, #3, #4, #5}
\def\book #1;#2;#3{\par\pp #1 #2, #3}
\def\rep #1;#2;#3{\par\pp #1 #2, #3}

\def\undertext#1{$\underline{\smash{\hbox{#1}}}$}
\def\simlt{\lower.5ex\hbox{$\; \buildrel < \over \sim \;$}}
\def\simgt{\lower.5ex\hbox{$\; \buildrel > \over \sim \;$}}

\def\etal{{et~al.}}
\def\noi{\noindent}
\def\bs{\bigskip}
\def\ms{\medskip}
\def\ss{\smallskip}
\def\ob{\obeylines}
\def\l{\line}
\def\hrf{\hrulefill}
\def\hf{\hfil}
\def\q{\quad}
\def\qq{\qquad}
\renewcommand{\deg}{$^{\circ}$}
\newcommand{\um}{$\mu$m}
\newcommand{\uk}{$\mu$K}
\newcommand{\qrms}{$Q_{rms-PS}$}
\newcommand{\n}{$n$}
\newcommand{\cdmr}{${\bf c}_{\rm DMR}$}
\newcommand{\xrms}{$\otimes_{RMS}$}
\newcommand{\gt}{$>$}
\newcommand{\lt}{$<$}
\newcommand{\ldl}{$< \delta <$}


\bs
\bs
\bs
\bs
\bs
\bs
\bs
\bs
\title{Non-cosmological Signal Contributions to the 
$COBE$\altaffilmark{1}-DMR Four-Year Sky Maps}

\author{A.J. Banday\altaffilmark{2,3,4}, 
K.M. G\'orski\altaffilmark{2,5}, 
C.L. Bennett\altaffilmark{6},
G. Hinshaw\altaffilmark{2},
A. Kogut\altaffilmark{2} \& G.F. Smoot\altaffilmark{7}.
}
\noindent
\altaffiltext{1}{The National Aeronautics and Space Administration/Goddard 
Space Flight Center (NASA/GSFC) is responsible for the design, development, 
and operation of the {\it Cosmic Background Explorer (COBE)}. Scientific 
guidance is provided by the $COBE$ Science Working Group. GSFC is also 
responsible for the development of the analysis software and for the 
production of the mission data sets.}
\altaffiltext{2}{Hughes STX Corporation, 
		Laboratory for Astronomy and Solar Physics,
		Code 685, NASA/GSFC, Greenbelt MD 20771.}
\altaffiltext{3}{Current address: Max Planck Institut f\"{u}r Astrophysik,
		85740 Garching Bei Munchen, Germany.}
\altaffiltext{4}{e-mail: {\it banday@ceylon.gsfc.nasa.gov}}
\altaffiltext{5}{on leave from Warsaw University Observatory, 
                 Aleje Ujazdowskie 4, 00-478 Warszawa, Poland.}
\altaffiltext{6}{Laboratory for Astronomy and Solar Physics,
		Code 685, NASA/GSFC, Greenbelt MD 20771.}
\altaffiltext{7}{LBL, SSL, \& CfPA, Bldg 50-25, University of California, 
                 Berkeley CA 94720.}

\bs
\bs
\bs

\begin{abstract}

We limit the possible contributions from non-cosmological sources to the
$COBE$-DMR four-year sky maps. The DMR data are cross-correlated with maps of 
rich clusters, extragalactic $IRAS$ sources, $HEAO-1$ A-2 X-ray
emission and 5 GHz radio sources using a Fourier space technique.
There is no evidence of significant contamination 
by such sources at an rms level of $\sim$ 8 \uk\ (95\% confidence level
at 7\deg\ resolution)
in the most sensitive 53 GHz sky map. This level is consistent with 
previous limits set by analysis of earlier DMR data and by simple extrapolations
from existing source models.
We place a 95\% confidence level rms limit on the Comptonization
parameter averaged over the high-latitude sky of
$\delta y\ <\ 1\ \times\ 10^{-6}$. Extragalactic sources
have an insignificant effect on CMB power spectrum parameterizations
determined from the DMR data.

\end{abstract}


\section{INTRODUCTION}

In this {\it Letter} 
we address the potential for contamination of the cosmic microwave background 
(CMB) anisotropy in the $COBE$-DMR four-year sky maps
by astrophysical foregrounds outside of our Galaxy.
Following Bennett et al.\ (1992),
we investigate possible
contributions from non-cosmological sources 
by direct comparison of the sky maps with
similar maps generated from extragalactic surveys. 

Models of the point source contribution at the DMR frequencies 
(Toffolati \etal, 1995) suggest a contribution 
$\Delta T / T < 10^{-7}$, well below the
level at which they may be of concern. 
Simple frequency extrapolations of the
$IRAS$ point source fluxes (Gawiser \& Smoot, 1996) reach the same conclusion.
However, such calculations are not sensitive to the more extended diffuse
emission which some of the maps considered here are. 

\section{METHOD}

Inverse-noise-variance weighted combinations of the A and B channels in Galactic pixelization
are formed at
each of the DMR frequencies, the weights being (0.618, 0.382), (0.579, 0.421)
and (0.382, 0.618) at 31.5, 53 and 90 GHz respectively. 
We cross-correlate the
DMR maps with a given extragalactic sky map $X$ over the high latitude sky,
defined as a region with Galactic latitude $\mid b\mid\ >$ 20\deg\ 
with additional cut sky regions at Ophiuchus and Orion 
(Bennett et al.\ 1996),
and at specific positions in the extragalactic maps with 
missing or unusable data. 
Each map is then decomposed into Fourier
coefficients using a basis of orthonormal functions explicitly computed 
on the available sky coverage (G\'orski 1994).
This technique has the advantage of allowing exact exclusion 
of the physically unconstrained monopole and dipole components 
from the analysis, and makes full use of all of the avilable 
spatial (phase) information.
The measured DMR
map Fourier coefficients, \cdmr, can then be written in vector form as
${\bf c}_{\rm DMR}\ = {\bf c}_{\rm CMB}\ +\ {\bf c}_{\rm N}\ 
+\ \alpha_X {\bf c}_{X}$ where ${\bf c}_{\rm CMB}$, ${\bf c}_{\rm N}$
and ${\bf c}_{X}$ are the coefficients describing the true CMB distribution,
the noise and the extragalactic map. $\alpha_X$ is now a coupling constant
(with units \uk\ $X^{-1}$) to be determined by minimizing 

\[
\chi^{2}\ =\ ({\bf c}_{\rm DMR}\ -\ \alpha_X {\bf c}_{X})^{T}\, {\bf\rm \tilde{M}}^{-1}\,
             ({\bf c}_{\rm DMR}\ -\ \alpha_X {\bf c}_{X})
\]
where {\bf \~{M}} is the covariance matrix describing the correlation
between different Fourier modes on the cut-sky and is dependent on
assumed CMB model parameters and the DMR instrument noise. The CMB
may be described by a Harrison-Zel'dovich power law spectrum
with rms quadrupole normalisation \qrms\ = 18.5 \uk. 
The Fourier coefficient vectors
contain spectral information over a range [$\ell_{min}$, $\ell_{max}$],
where $\ell_{min}$ is either 2 (quadrupole included) 
or 3 (quadrupole excluded). 
We set $\ell_{max}=30$, which is sufficient
to describe the cosmic content of the DMR data.

The coupling constant $\alpha$ has an exact solution
\[ 
\alpha_X\ =\ {\bf c}_{X}\, {\bf\rm \tilde{M}}^{-1}\, {\bf c}_{\rm DMR}/
           {\bf c}_{X}\, {\bf\rm \tilde{M}}^{-1}\, {\bf c}_{X}
\]
and a Gaussian error given by $\sigma^{2}(\alpha_X)$ =\
$({\bf c}_{X}\, {\bf\rm \tilde{M}}^{-1}\, {\bf c}_{X})^{-1}$.

The contribution of $X$ to the DMR sky is the map $\alpha_X T_X$,
which can be characterized 
by the rms amplitude $(\alpha_X T_X)_{rms}$ and peak-to-peak
quantity $(\alpha_X T_X)_{pp}$. Both values are computed over the
cut-sky after subtraction of a best fit monopole, dipole and, if appropriate,
quadrupole.

\section{ABELL-CORWIN-OLOWIN RICH CLUSTERS}

Inverse Compton scattering of CMB photons from hot intracluster gas causes
a decrement in the Rayleigh-Jeans part of the CMB spectrum, 
$\Delta T\ =\ -2yT_0$, where $T_0$ is the unperturbed CMB temperature
and $y$, the Comptonization parameter, is a function of the integrated
electron pressure along the line-of-sight. This Sunyaev-Zel'dovich effect
(Sunyaev \& Zel'dovich, 1980) may generate some of the temperature
anisotropy observed by $COBE$-DMR. As in Bennett \etal\ (1993), we
use a rich cluster map constructed from the data of Abell, Corwin \& Olowin
(1989, hereafter ACO) to place limits on this contribution. We therefore assume
that the ACO rich cluster map is a good tracer of the spatial morphology
of the large scale cluster gas distribution giving rise to the SZ effect,
and that the richness class is a good indicator of the optical depth
of the gas. Each pixel of the rich cluster map is a sum over the
richness classes of all clusters in that pixel. 
The map is convolved with the DMR beam pattern (E.L.\ Wright \etal, 1994).
Additional versions of the map have been computed with different Abell
richness thresholds, and by summing over different exponents of the 
richness class in a given pixel. No significant differences are seen for
the different binning strategies. We report results for a minimum
richness class threshold of 1, and simple binning by the richness class
(ie. an exponent of 1).

Table 1 summarizes the results of the analysis. 
The SZ effect will generate approximately 
equal and negative $\alpha_X$ values over the three DMR frequencies.
There are no significant correlations
between the DMR sky maps and the ACO rich cluster map. The 31 GHz 
coupling coefficient is negative, though not at a statistically significat level.
However, it is more likely
due to anti-correlation between the ACO zone of avoidance and the DMR galactic
signal. This is supported by the decrease in significance for the quadrupole
excluded fit, which would remove a significant fraction of 
the Galactic emission. 

\section{{\it HEAO-1} A-2 X-RAY DATA}

The $HEAO-1$ A-2 X-ray emission map, as discussed in Bennett \etal\ (1993) 
and provided by K. Jahoda with flux scale as in Jahoda \& Mushotzky (1989),
was reconvolved to the DMR beam profile and cross-correlated with the DMR
sky maps. With the DMR custom Galactic cut, 3875 pixels remain
at high latitude. Strong Galactic emission from the Ophiuchus and Orion regions
common to both the X-ray and DMR maps is excised by this process. 
Nevertheless, there remain correlations in excess of $2\, \sigma$
which may be associated with either the LMC (a prominent high latitude
feature in the $HEAO-1$ map) or perhaps from Galactic emission
traced by the X-ray map. 
At 95\% confidence we limit the contribution of anisotropy
traced by the X-ray map to the DMR analysis to be $<$ 8 \uk\,
irrespective of its origin. 

Since rich clusters of galaxies are bright X-ray 
sources the X-ray emission in the $HEAO-1$ map could also, in principle,
trace the hot gases in these structures which give rise to the SZ effect
(although the angular distribution of the X-ray emission and temperature
anisotropy would be somewhat different).
We might then expect some negative correlation with the DMR maps (since at DMR
frequencies the SZ effect manifests itself as a temperature decrement), but in
all cases the correlation constants are positive. 
We conclude, like Bennett \etal\ (1993), 
Boughn \&\ Jahoda (1993), and
Rephaeli (1993),
that the fluctuations observed by $COBE$ are not
generated by such a mechanism.
A chance alignment of features of primordial CMB anisotropy with features
in the X-ray map would give similar flat-spectrum correlations with $\sim$ 3\% probability.

\section{1.2 JANSKY {\it IRAS} GALAXIES}

We consider a sample of $IRAS$ galaxies selected from the $IRAS$ point 
source catalog for which redshifts have been measured.
Fisher \etal\ (1995) extend the 2 Jy survey of Strauss \etal\ (1992) to a flux
limit of 1.2 Jy, resulting in a catalog of some 5321 galaxies with measured
redshifts covering 87.6\% of the sky. A map of the 1.2 Jy sky was
generated as described previously using the $IRAS$ 100 $\mu$m
flux, but excluding regions as prescribed in Strauss \etal\ (1990) and
using a mask and software provided by Strauss (1995).
The $IRAS$ 100 $\mu$m sky is dominated by emission from the Large 
Magellanic Cloud (LMC) but we are more interested in potentially significant
correlations with the rest of the map. Fortunately, the mask does exclude
a region around the LMC. After applying the mask and the custom DMR
galactic cut 3620 pixels remain. 
The $IRAS$ galaxies show notable structure around Virgo, Coma
and in the supergalactic plane; however, no correlations exceed $\sim$ 
$2\, \sigma$ significance.

\section{5 GHZ RADIO SOURCES}

We have combined data from the 87 GB 5 GHz survey of Gregory \& Condon (1991)
with new southern hemisphere data from the Parkes-MIT-NRAO (PMN) surveys. 
The PMN data are divided into declination ranges as follows:
the southern survey covers -87\deg .5 \ldl -37\deg (A.E. Wright \etal, 1994),
the zenith -37\deg \ldl -29\deg (Tasker \etal, 1996),
the tropical -29\deg \ldl -9\deg .5 (Griffith \etal, 1994)
and the equatorial -9\deg .5 \ldl +10\deg\ (Griffith \etal, 1995).
Together with the 87 GB coverage of 0\deg \ldl +75\deg, we should
in principle be able to construct a nearly all-sky radio source map.
Unfortunately, the various data sets are complete at different flux limits,
ranging from $\sim$ 25 mJy for the 87 GB data and parts of the southern 
survey to $\sim$ 72 mJy for the zenith survey (A.E.\ Wright, 1995 -- private 
communication). Two maps were therefore assembled: the first with a lower flux
threshold of 40 mJy and $excluding$ the zenith survey data, the second with
a threshold of 72 mJy and $including$ the zenith survey. In both cases, 
data in the declination range 0\deg \ldl 10\deg were taken from the 87 GB
survey: although some inconsistencies exist between the 87 GB and PMN 
catalogs over this common declination band (Gregory \etal, 1994), small
positional uncertainties are unlikely to be of importance after
smoothing to DMR resolution. Large systematic differences in the
flux densities would generate a strong gradient between the northern and southern
celestial hemispheres which is not observed in the maps\footnote{Furthermore,
replacing the PMN southern survey
with an alternative version from Gregory \etal\ (1994) generated in
a similar fashion to the 87 GB catalog also yields statistically 
consistent results.}. 

Cross-correlation of both maps
with the DMR data shows no statistically significant difference in the results,
and we have elected to continue the analysis with the 72 mJy flux limited
map which has the greater sky coverage ($\sim$ 12 steradians).
Off the Galactic plane much of the signal appears towards Virgo.
After applying the extended galaxy cut to the map 3690 usable pixels 
(covering $\sim$ 7.5 steradians) remain. No correlations exceed the 
$\sim$ $1\, \sigma$ level.

\section{JOINT ANALYSIS}

We have extended the Fourier space template fitting technique as described
in G\'orski \etal\ (1996) to enable us to fit the four non-cosmological
template maps to the three DMR frequencies simultaneously. 
This forces the CMB distribution (in thermodynamic units)
to be invariant between the three frequencies, thus
allowing an improved determination of the extent of chance
cosmic or noise alignments with the template maps.
We include the
quadrupole components in the analysis, and determine the galactic
contribution by fitting to two Galactic templates -- the $DIRBE$ 140 \um\ map
and a radio survey at 408 MHz (Haslam \etal, 1981).
Kogut \etal\ (1996a) have discussed the effectiveness
of these two maps in accounting for Galactic dust, free-free and synchrotron 
emission, and the Galactic microwave emission at high latitudes has
been reconsidered in the light of the DMR four-year results in Kogut \etal\
(1996c). 
Orthonormal functions
were computed for the sky coverage common to all template maps, 
leaving 3423 pixels after the custom DMR Galactic cut excision (the
region around the LMC was therefore masked since it is removed in the $IRAS$ 
1.2 Jy analysis). 
No constraints were imposed on the spectral dependences of the template maps
over the DMR frequencies.
Since the method considers 
the three DMR maps simultaneously, it results in the strongest
constraints on non-cosmological correlations.
18 coefficients were determined, 
the non-Galactic 12 of which are summarized
in column (8) of Table 1. The results are in excellent agreement
with the single DMR channel-to-template fits. 

The correlation between
the 31 GHz map and the ACO Rich Clusters template becomes positive,
following the trend evident from the fits with and without quadrupole.
This supports the previous conclusion that the initially negative 
values for the $\alpha_X$ coefficient are due to a chance 
alignment of features rather than any genuine
SZ contribution. Although there is some cross-talk between the
31 GHz cluster and $DIRBE$ correlations, this is not unexpected since
the ACO zone of avoidance anti-correlates with the 
predominantly quadrupolar Galactic signal.
Typical uncertainties in $\alpha$ give a 95\% confidence level upper limit 
on the 7\deg\ rms contribution traced by the ACO map of better than
$\sim$ 6 \uk\ for the sensitive 53 and 90 GHz channels. This yields
a corresponding 95\% CL rms limit on the Comptonization parameter 
averaged over the high
latitude sky of $\delta y\ =\ -(\delta T/2T)\ \simeq\ 1\ \times\ 10^{-6}$.
If we interpret the limits from the $HEAO-1$ A-2 correlations
similarly (i.e. as limits on SZ related fluctuations), this limit improves
to better than $8.4\ \times\ 10^{-7}$. Rephaeli (1993) has independently
searched the $HEAO-1$ A-2 data base to place direct bounds on the properties
of hot gas in superclusters of galaxies, placing a limit on the mean
Comptonization parameter of a few $\times\ 10^{-7}$,
in excellent agreement with the results here.

In fact,
the correlations between the DMR maps
and both the ACO clusters and $HEAO-1$ A2 X-ray map
are predominantly positive, in contrast to the expected signature
from the SZ effect. Such a signal could be indicative
of radio sources embedded in the clusters. Since the cross-correlation
between these maps and the 5 GHz radio sources is negligible,
this would require that the dominant sources at DMR frequencies had
flat spectra and were below the detection threshold at the lower frequency.
We place a conservative 95\% CL upper limit 
on the rms $emission$ from clusters over the high latitude sky
of 10 \uk. 

Note that, with the exclusion of the LMC region
from the joint analysis sky coverage, the $2.5\, \sigma$ correlation
between the 53 GHz sky map and the $HEAO-1$ A2 X-ray map is
significantly decreased. This is also the case when a single map
correlation is computed between the 53 GHz and X-ray maps after removing
a circular region of 10\deg\ radius on the LMC (for which $\alpha$ falls
to $\sim 20\pm 15$). A significant fraction
of the correlation appears to be associated with this region.
Bennett \etal\ (1993) predicted that the LMC, and specifically the giant HII 
region 30 Doradus, might contribute a signal of order 50 \uk\ at 53 GHz.
Kogut \etal\ (1994) determined an upper limit of 87 \uk\ (95\% confidence level)
from the two year sky maps with
a point source fitting technique, in good agreement with this prediction.
Using the peak-peak limits from the earlier fits including this
region, we place 95\% upper limits to the antenna temperature
microwave emission of the LMC of 109, 49 and 57 \uk\ at 31.5, 53
and 90 GHz respectively. 

In addition, the correlation coefficients between the Galactic templates
and the 4-year maps are in good agreement with those determined
in Kogut \etal\ (1996c). For comparison, at 53 GHz, we find 
$\alpha_{140 \mu {\rm m}}$ = $2.45\pm 1.04$, 
$\alpha_{408 {\rm MHz}}$ = $0.36\pm 0.76$ to convert the Galactic templates
to antenna temperature in \uk. There is no significant cross-talk between the 
Galactic and extragalactic templates.

Finally, we have tested the sensitivity of the fits to the assumption
that the cosmic signal present in the DMR sky maps is well
described by a power law spectrum with \qrms\ = 18.5 \uk\
and \n\ = 1 by running a full grid of power law models as 
in G\'orski \etal\ (1996).
After weighting the $\alpha$ coefficients by the 
likelihood function determined
for the grid of cosmological models, 
we find that the Harrison-Zel'dovich model computations are
perfectly adequate. Moreover, we find that
the effect of the non-cosmological template subtraction
on the most likely (\qrms, \n) is to shift these values
by $\sim$0.1 \uk\ and -0.06 respectively.
We conclude that there is no evidence for significant non-cosmological
signal contamination of the DMR sky maps.

\section{CORRELATION WITH INDIVIDUAL BRIGHT CLUSTERS}

In the previous sections, we have demonstrated that there is
no evidence of non-cosmological signal detection based on correlation
of the DMR maps with fixed source signal templates. This limit 
applies to the spatial average of sources at high Galactic latitudes,
and so we consider here whether individual bright clusters of galaxies
are detected in the DMR data. Nearby clusters are more likely to contribute
a measurable signal in the DMR sky maps given their large angular extent.
We consider the mean temperatures in the DMR maps towards 5 nearby clusters
to search for indications of SZ decrements or radio and
dust emission from galaxies within the cluster (which are competing
effects at the DMR frequencies, and any limits should be considered
in light of this). A \lq\lq ring" technique,
as utilized previously in Bennett \etal\ (1993) and Kogut \etal\ (1994),
computes the difference between the mean temperature of pixels within
a disk of radius $\theta$ and those in the surrounding annulus of equal area.
This technique should be sensitive to low surface Comptonization. The results
summarized in Table 2 are for $\theta\ =\ 10^{\circ}$ . 
There are no significant detections when a weighted mean
is taken over the three DMR frequencies. An upper limit of 38 \uk\ can
be placed on emission from nearby clusters. 

Herbig \etal\ (1995) have detected
the SZ effect in Coma with a central decrement of $-505\pm 92$ \uk\ (after 
large corrections for discrete radio sources and applying a model for the X-ray
atmosphere of the cluster). It is not surprising that DMR is unable
to detect this anisotropy given the large beam dilution. The Coma cluster
has an approximate 0\deg .5 FWHM, which by virtue of its ratio
to the DMR beam solid angle implies an expected signal of less than 3 \uk.
A limit of $\delta y\ < 7\ \times\ 10^{-6}$
can be established for Comptonization by nearby clusters (from the maximum
weighted error in Table 2).

\section{DISCUSSION AND CONCLUSIONS}

We have examined the $COBE$ DMR 4-year sky maps for evidence of non-cosmological
signal contamination and find none. A simultaneous fit of the three DMR 
frequencies to two Galactic templates and four extragalactic 
sky maps yields rms limits due to the extragalactic correlations alone
of $\sim$ 18, 8 and 11 \uk\ at 31.5, 53 and 90 GHz
respectively (95\% confidence levels). These limits are higher than those predicted
by simple point source models, but are limited by the DMR noise levels
and might be regarded as somewhat conservative. 
The contribution from non-cosmological sources is insubstantial 
compared to the $\sim$ 35 \uk\ rms anisotropy signal detected in the DMR sky maps
at 7\deg\ resolution (Banday et al.\ 1996). 
Moreover, 
extragalactic template corrections perturb the maximum likelihood power 
spectrum analysis of the $COBE$-DMR data for power law cosmological models
insignificantly: the best fit parameters (\qrms, \n) are only shifted 
by $\sim$0.1 \uk\ and -0.06 respectively from their uncorrected values.

We acknowledge the efforts of those contributing to the $COBE$ DMR.
$COBE$ is supported by the Office of Space Sciences of NASA Headquarters.
We thank Charley Lineweaver and Luis Tenorio for useful comments.

\clearpage

{\tiny

\begin{table}[h]
{\small
\noindent Table 1: Cross-correlation results and non-cosmological signal 
contributions.}
\begin{center}

\vspace{2mm}

\begin{tabular}{cccccccccc}\hline\hline
 & \multicolumn{3}{c}{quadrupole included} & \multicolumn{3}{c}{quadrupole
excluded} & 
\multicolumn{3}{c}{joint analysis} \\
Freq & $\alpha_X$ & ($\alpha T_X)_{rms}$ & ($\alpha T_X)_{pp}$ &
       $\alpha_X$ & ($\alpha T_X)_{rms}$ & ($\alpha T_X)_{pp}$ &
       $\alpha_X$ & ($\alpha T_X)_{rms}$ & ($\alpha T_X)_{pp}$ \\ 
(GHz) & ($\mu$K $X^{-1}$) & ($\mu$K) & ($\mu$K) &
        ($\mu$K $X^{-1}$) & ($\mu$K) & ($\mu$K) &
        ($\mu$K $X^{-1}$) & ($\mu$K) & ($\mu$K) \\ \hline
\multicolumn{10}{l}{\bf ACO Rich Clusters} \\ 
31 & $-10.34\pm 8.74$ & $6.2\pm 5.3$ & $47.9\pm 39.9$ &  
     $-4.60\pm 9.64$ & $2.1\pm 4.4$ & $17.9\pm 37.4$ &   
     $0.37\pm8.51$ & $0.2\pm 5.0$ & $1.3\pm 30.7$ \\      
53 & $1.72\pm 4.78$ &   $1.0\pm 2.9$ & $7.8\pm 21.8$ &
     $2.79\pm 4.97$ & $1.3\pm 2.3$ & $10.8\pm 19.3$ &
     $5.85\pm 4.52$ & $3.4\pm 2.6$ & $21.1\pm 16.3$ \\
90 & $3.30\pm 6.04$ &   $2.0\pm 3.6$ & $15.1\pm 27.5$ &
     $3.77\pm 6.38$ & $1.7\pm 2.9$ & $14.6\pm 24.8$ &
     $1.10\pm 5.52$ & $0.6\pm 3.2$ & $4.0\pm 19.9$ \\
\multicolumn{10}{l}{\bf $HEAO-1$ A2 X-ray sources} \\
31 & $46.60\pm 23.52$ & $7.5\pm 3.8$ & $112.5\pm 56.8$ &
     $43.71\pm 23.69$ & $6.8\pm 3.7$ & $102.8\pm 55.7$ &
     $52.87\pm 26.79$ & $7.6\pm 3.8$ & $113.2\pm 57.35$ \\
53 & $28.12\pm 11.09$ & $4.6\pm 1.8$ & $67.9\pm 26.8$ &
     $27.97\pm 11.10$ & $4.4\pm 1.7$ & $65.8\pm 26.1$ &
     $4.65\pm 12.14$ & $0.7\pm 1.7$ & $10.0\pm 26.0$ \\
90 & $22.84\pm 14.93$ & $3.7\pm 2.4$ & $55.1\pm 36.0$ &
     $22.89\pm 14.97$ & $3.6\pm 2.3$ & $53.8\pm 35.2$ &
     $15.54\pm 16.21$ & $2.5\pm 2.3$ & $37.5\pm 34.7$ \\
\multicolumn{10}{l}{\bf 1.2 Jy $IRAS$ Galaxies} \\
31 & $0.75\pm 0.42$ & $8.0\pm 4.5$ & $124.1\pm69.3$ &
     $0.85\pm 0.42$ & $8.7\pm 4.3$ & $138.8\pm 68.6$ &  
     $0.48\pm 0.37$ & $5.3\pm 4.1$ & $80.3\pm 61.9$ \\
53 & $0.21\pm 0.19$ & $2.3\pm 2.1$ & $35.2\pm 32.4$ &
     $0.22\pm 0.20$ & $2.3\pm 2.0$ & $36.3\pm 31.7$ &
     $0.04\pm 0.17$ & $0.5\pm 1.9$ & $7.4\pm 28.3$ \\
90 & $0.01\pm 0.26$ & $0.2\pm 2.8$ & $2.5\pm 43.3$ &
     $0.02\pm 0.26$ & $0.2\pm 2.7$ & $2.8\pm 42.6$ &
     $0.27\pm 0.22$ & $3.0\pm 2.4$ & $45.0\pm 37.1$ \\
\multicolumn{10}{l}{\bf 5 GHz (87 GB/PMN) Radio Sources} \\
31 & $0.09\pm 0.14$ & $2.9\pm 4.4$ & $69.9\pm 106.8$ &
     $0.08\pm 0.15$ & $2.5\pm 4.4$ & $60.7\pm 105.7$ &
     $-0.09\pm 0.15$ & $2.7\pm 4.6$ & $65.4\pm 111.0$ \\
53 & $0.09\pm 0.06$ & $2.6\pm 1.9$ & $62.7\pm 45.9$ &
     $0.09\pm 0.06$ & $2.6\pm 1.9$ & $62.5\pm 45.3$ &
     $-0.01\pm 0.06$ & $0.4\pm 1.9$ & $9.4\pm 47.0$ \\
90 & $0.11\pm 0.09$ & $3.4\pm 2.6$ & $81.7\pm 62.9$ &
     $0.11\pm 0.09$ & $3.4\pm 2.6$ & $81.8\pm 62.1$ &
     $0.06\pm 0.09$ & $1.7\pm 2.6$ & $40.9\pm 63.0$ \\ \hline

\end{tabular}
\end{center}

\end{table}

{\small
\noindent $Note$ -- rms and peak-peak values are computed after
subtraction of the monopole, dipole and, where appropriate, quadrupole
distributions over the uncut regions of the sky.
All temperatures are in thermodynamic units.}

}

\clearpage

\begin{table}[h]
\noindent Table 2: Fitted amplitudes toward Sunyaev-Zel'dovich candidate
clusters (\uk).

\begin{center}

\vspace{2mm}

\begin{tabular}{ccccc} \hline\hline
Source & $T_{31}$ & $T_{53}$ & $T_{90}$ & Weighted Mean\\ \hline
\multicolumn{1}{l}{Coma} & $-3\pm 52$ & $4\pm 16$ & $ -59\pm 25$ & $-14\pm 13$ \\
                         & $-63\pm 53$ & $5\pm 16$ & $20\pm 26$ & $5\pm 13$ \\
\multicolumn{1}{l}{Virgo} & $54\pm 59$ & $-21\pm 17$ & $33\pm 26$ & $-2\pm 14$ \\
                          & $-35\pm 60$ & $10\pm 17$ & $-33\pm 27$ & $-4\pm 14$ \\
\multicolumn{1}{l}{Perseus-Pisces} & $16\pm 67$ & $1\pm 24$ & $3\pm37$ & $3\pm 19$ \\
                                   & $-32\pm 69$ & $25\pm 24$ & $4\pm 38$ & $15\pm 19$ \\
\multicolumn{1}{l}{Hercules} & $105\pm 51$ & $9\pm 19$ & $-3\pm 30$ & $15\pm 15$  \\
                             & $-54\pm 52$ & $13\pm 19$ & $15\pm 31$ & $8\pm 15$ \\
\multicolumn{1}{l}{Hydra} & $-112\pm 62$ & $-29\pm 21$ & $9\pm 32$ & $-25\pm 17$  \\
                          & $77\pm 64$ & $-18\pm 21$ & $12\pm 33$ & $-3\pm 17$  \\ \hline

\end{tabular}
\end{center}

\end{table}

\noindent $Note$ -- First line for each source refers to fitted amplitude of
weighted sum map, the second refers to the (A-B)/2 difference map. All 
amplitudes are thermodynamic temperature units.


\begin{references}

\ref Abell, G.O., Corwin, H.G. \& Olowin, R.P.;1989;ApJS;70;1

\rep Banday, A.J., G\'orski, K.M., Bennett, C.L., Hinshaw, G., Kogut, A. 
Lineweaver, C., Smoot, G.F. \& Tenorio, L.;1996;ApJ submitted

\ref Bennett, C.L., Hinshaw, G., Banday, A., Kogut, A., Wright, E.L., 
Loewenstein, K. \& Cheng, E.S.;1993;ApJ;414;L77

\rep Bennett, C.L., et al.;1996;ApJ submitted




\ref Boughn, S.P. \& Jahoda, K.;1993;ApJ;412;L1

\ref Fisher, K.B., Huchra, J.P., Strauss, M.A., Davis, M., Yahil, A. \&
Schlegel, D.;1995;ApJS;100;69

\rep Gawiser, E. \& Smoot, G.F.;1996; in preparation

\rep G\'orski, K.M., Banday, A.J., Bennett, C.L., Hinshaw, G., Kogut, A.,
Smoot, G.F., \& Wright, E.L.;1996; ApJ submitted

\ref G\'orski, K.M.;1994;ApJ;430;L85


\ref Gregory, P.C. \& Condon, J.J.;1991;ApJS;75;1011

\ref Gregory, P.C. Vavasour, J.D., Scott, W.K. \& Condon, J.J.;1994;ApJS;90;173

\ref Griffith, M.R., Wright, A.E., Burke, B.F., \& Ekers, R.D.;1994;ApJS;90;179

\ref Griffith, M.R., Wright, A.E., Burke, B.F., \& Ekers, R.D.;1995;ApJS;97;347

\ref Haslam, C.G.T., Klein, U., Salter, C.j., Stoffel, H., Wilson, W.E., 
Cleary, M.N., Cooke, D.J. \& Thomasson, P.;1981;A\&A;100;209

\ref Herbig, T., Lawrence, C.R., Readhead, A.C.S. \& Gulkis, S.;1995;ApJ;449;L5

\ref Hogan, C.;1992;ApJ;398;L77

\rep Hinshaw, G., Banday, A.J., Bennett, C.L., G\'{o}rski, K.M.,
Kogut, A., Smoot, G.F., \& Wright, E.L.;1996;ApJ submitted

\ref Jahoda, K. \& Mushotzky, R.F.;1989;ApJ;346;638


\ref Kogut, A., Banday, A.J., Bennett, C.L., Hinshaw, G., Loewenstein, K.,
Lubin, P., Smoot, G.F. \& Wright, E.L.;1994;ApJ;433;435

\rep Kogut, A., Banday, A.J., Bennett, C.L., G\'orski, K.M., Hinshaw, G. \&
Reach, W.;1996a;ApJ 460 in press 	

\rep Kogut, A., et al.;1996b;ApJ submitted  

\rep Kogut, A., Hinshaw, G., Banday, A.J., Bennett, C.L., G\'orski, K.M.,
Smoot, G.F., \& Wright, E.L.;1996c;ApJ submitted    

\rep Kogut, A., Banday, A.J., Bennett, C.L., G\'orski, K.M., Hinshaw, G.,
Smoot, G.F., \& Wright, E.L.;1996d;ApJ submitted  

\rep Reach, W.T., Franz, B.A., Kelsall, T. \& Weiland, J.L.;1995;
Unveiling the Cosmic Infrared Background, ed. E. Dwek, (New York:AIP)


\ref Rephaeli, Y.;1993;ApJ;418;1

\ref Smoot, G.F., \etal;1992;ApJ;396;L1


\ref Strauss, M., Davis, M., Yahil, A., \& Huchra, J.P.;1990;ApJ;361;49

\ref Strauss, M., Huchra, J.P., Davis, M., Yahil, A., Fisher, K.B. \& 
Tonry, J.;1992;ApJS;83;29

\rep Strauss, M.;1995;NASA ADS supplement

\ref Sunyaev, R.A. \& Zel'dovich, Ya.B.;1980;ARA\& A;18;537

\rep Tasker, N., Wright, A.E., Griffith, M.R., \& Condon, J.J.;1995;AJ in press

\rep Toffolatti, L. \etal;1995;Astrophys. Letts. \& Comm. in press

\ref Wright, A.E., Griffith, M.R., Burke, B.F., \& Ekers, R.D.;1994;ApJS;91;111

\rep Wright, A.E.;1995;private communication


\ref Wright, E.L., Smoot, G.F., Kogut, A., Hinshaw, G., Tenorio, L.,
Lineweaver, C., Bennett, C.L. \& Lubin, P.M.;1994;ApJ;420;1

\rep Wright, E.L., Bennett, C.L., G\'orski, K.M., Hinshaw, G., 
\& Smoot, G.F.;1996;ApJ submitted


\end{references}
\end{document}